%% file: paper.tex
\documentclass[submission, Phys]{SciPost}

\pdfoutput=1
\input{incl_settings.tex}
\input{incl_shortcuts.tex}
\graphicspath{{./figures/}}

\begin{document}

\begin{center} 
    {\Large\textbf{\color{scipostdeepblue}{Know What You Don't Flow}}}
\end{center}

\begin{center}\textbf{
    Anja~Butter\textsuperscript{1,2},
    Sascha~Diefenbacher\textsuperscript{1},
    Tilman~Plehn\textsuperscript{1,3}, and
    Lorenz~Vogel\textsuperscript{1}
}\end{center}

\begin{center}
    {\bf 1} Institut f\"{u}r Theoretische Physik, Universit\"{a}t Heidelberg, Germany\\
    {\bf 2} LPNHE, Sorbonne Université, Université Paris Cité, CNRS/IN2P3, Paris, France\\
    {\bf 3} Interdisciplinary Center for Scientific Computing (IWR), Universit\"{a}t Heidelberg, Germany\\
\end{center}

\begin{center}
    \today
\end{center}
 
\section*{\color{scipostdeepblue}{Abstract}}
\textbf{\boldmath{Calibrated learned uncertainties are a key requirement also for generative neural networks in LHC physics. For a toy model with an explicit likelihood we show how a heteroscedastic and a Bayesian normalizing flow learn the systematic and statistical uncertainties on the underlying phase space density. Without an explicit likelihood we train the heteroscedastic loss on a classifier-reweighted approximate generative network. We illustrate our comprehensive approach for top pair events and show how a conditional heteroscedastic flow propagates calibrated uncertainties to all phase space directions.}}

\vspace{10pt}
\noindent\rule{\textwidth}{1pt}
\tableofcontents\thispagestyle{fancy}
\noindent\rule{\textwidth}{1pt}
\vspace{10pt}

\clearpage
\section{Introduction}
\label{sec:intro}

The development of the LHC into the first precision-hadron collider combines precisely measured kinematic distributions with equally precise quantum field theory predictions. In view of the high-luminosity LHC, the precision of the theory predictions has to match the increased statistical and systematic precision of the measurements. To reach this goal, generative machine learning offers a transformative methodology~\cite{Butter:2022rso,Plehn:2022ftl,Ubiali:2026myh,Kramer:2026eoq}. Generative adversarial networks~\cite{Otten:2019hhl,Hashemi:2019fkn,DiSipio:2019imz,Butter:2019cae,Alanazi:2020klf,Choi:2021sku}, normalizing flows~\cite{Gao:2020zvv,Stienen:2020gns,Bellagente:2021yyh,Butter:2021csz,Verheyen:2022tov}, diffusion networks, and transformer architectures~\cite{Butter:2023fov,Brehmer:2024yqw,Favaro:2025pgz,Petitjean:2025zjf} extract the phase-space density and generate new events extremely faster than the classical chain. The same generative network architectures can be used for neural importance sampling~\cite{Bendavid:2017zhk,Gao:2020vdv,Deutschmann:2024lml}, implemented in \madnis~\cite{Heimel:2022wyj,Heimel:2023ngj,DeCrescenzo:2026tsp,Heimel:2026cxh} and \sherpa~\cite{Gao:2020zvv, Bothmann:2020ywa, Bothmann:2025lwg,Janssen:2025zke}. Because the phase space density is assumed to be smooth, the generated sample can be closer to the truth than the finite-size training sample~\cite{Butter:2020qhk,Bieringer:2022cbs,Bieringer:2024nbc,Bahl:2025ryd,Bahl:2026xpz}. 

An alternative strategy only accelerates the expensive modules of the simulation chain, replacing the explicit matrix element by an ultra-fast regression network. Its main advantage is that we can complement the explicitly available phase space density with a learned uncertainty~\cite{Plehn:2022ftl,Haussmann:2026gbi}, using Bayesian neural networks~\cite{Kasieczka:2020vlh,Badger:2022hwf,Bahl:2024gyt}, heteroscedastic losses with Gaussian~\cite{Bahl:2024gyt,Beccatini:2025tpk} or other~\cite{Bahl:2026qaf} likelihoods, repulsive ensembles~\cite{Bahl:2024gyt}, evidential regression~\cite{Bahl:2025xvx}, the Donsker--Varadhan representation~\cite{Gambhir:2022gua}, conformal predictions~\cite{Araz:2025vuw,Dubey:2026cts}, or conditional nuisance parameter propagation~\cite{Butter:2021csz,Valsecchi:2026kjy}. For supervised amplitude regression we can learn statistical uncertainties, defined as vanishing with more of the same training data, separately from systematic uncertainties. The learned uncertainties are calibrated, which is important for LHC applications, for example, in parameter inference~\cite{Benato:2025rgo,Elsharkawy:2025six,Alvarez:2026umb}.

In this paper, we target a weak point of generative networks, the uncertainty quantification for implicitly learning an underlying phase space density from events. Bayesian normalizing flows~\cite{Bellagente:2021yyh,Butter:2021csz}, diffusion models~\cite{Butter:2023fov}, autoregressive transformers~\cite{Butter:2023fov,Butter:2024zbd}, and weighted ensembles~\cite{Benevedes:2025nzr} have shown that they can learn the (statistical) uncertainty reflecting a lack of training data in a given phase space region. However, there exist no particle physics studies that show the correct calibration of learned statistical uncertainties of generative networks. Moreover, systematic uncertainties on the learned density, for instance, from the network expressivity, have not been studied in any detail. This is why we tackle the combination of learned statistical and systematic uncertainties and their calibration in this paper, from a conceptual perspective and for a realistic LHC application.

We first develop a normalizing flow framework to learn different sources of uncertainties in \cref{sec:flow}. This includes a heteroscedastic normalizing flow (HNF) to learn the systematic uncertainty for a known density and a Bayesian normalizing flow (BNF) to learn the statistical uncertainty. For cases where we do not know the underlying density explicitly we propose to use a re-weighted initial estimate to train the final normalizing flow with an uncertainty-aware heteroscedastic loss. In \cref{sec:toy} we show how these concepts provide us with calibrated systematic and statistical uncertainties for a toy model. As our physics application, we look at leptonic top pair events in \cref{sec:physics}, where the underlying phase space density is not known explicitly. We show that the reweighting strategy with a heteroscedastic normalizing flow and the Bayesian normalizing flow indeed learn local calibrated uncertainties for LHC applications.

\clearpage
\section{Uncertainty-aware generation}
\label{sec:flow}

A proper uncertainty framework for generative networks requires us to distinguish different sources of uncertainties and test their calibration. Usually, generative networks are trained in an unsupervised manner, \ie, the training data do not provide the true density and instead encode it in unweighted events. We first employ a toy model for which the explicitly known true density allows for an uncertainty treatment analogous to heteroscedastic regression. For this study, we rely on normalizing flows, specifically the symmetrically invertible INN, but many of the conclusions can be generalized to other generative architectures.

\subsection{Heteroscedastic normalizing flow}
\label{sec:flow_het}

At the LHC, normalizing flows (NFs)~\cite{Gao:2020zvv,Stienen:2020gns,Verheyen:2022tov} and their symmetric invertible neural network (INN) variant~\cite{Bellagente:2021yyh,Butter:2021csz}, map a simple latent distribution to a phase space distribution encoded in the training data through a bijective transformation,
\begin{align}
    r\sim \pl(r)
    \quad\xleftrightarrow[\quad\leftarrow \overline{G}_{\theta}(x)\quad]{\quad G_{\theta}(r) \rightarrow\quad}\quad
    x\sim \pmodel(x) \stackrel{\text{train}}{\approx} \ptruth(x) \eqp
    \label{eq:bijection}
\end{align}
The phase space density $\pmodel(x)$ is given in terms of a learned Jacobian 
\begin{align}
    \pmodel(x)
    = \pl(\overline{G}_{\theta}(x)) \: \left| \frac{\partial \overline{G}_{\theta}(x)}{\partial x} \right| \eqp
    \label{eq:jacobian}
\end{align}
The latent distribution is usually a standard multivariate Gaussian or a uniform distribution,
\begin{align}
    \pl(r) = \gauss(0, \mathbb{1})
    \qquad\mor\qquad
    \pl(r) = \mathcal{U}([0, 1]) \eqp
    \label{eq:latent-uniform}
\end{align}
NFs are trained by minimizing the negative log-likelihood loss,
\begin{align}
    \loss_{\text{NF}}
    = -\log \pmodel(x)
    = -\log \pl(\overline{G}_{\theta}(x))
    - \log \left| \frac{\partial \overline{G}_{\theta}(x)}{\partial x} \right| \eqp
    \label{eq:loss-nf}
\end{align}
This likelihood training requires samples $x \sim \ptruth(x)$. 

Because the NF encodes the Jacobian and it is not obvious how a single network will learn the density and its uncertainty, we jointly train $(i)$ a NF, parameterized by $\theta$, predicting $\log\pmodel(x)$; and $(ii)$ an MLP encoding the uncertainty $\sigsyst(x)$ of $\pmodel(x)$. For an explicit target density, we sample random phase space points $\{x\}$, compute $\ptruth(x)$, and inverse-pass them to obtain $r = \overline{G}_{\theta}(x)$. This gives us the Jacobian determinant in Eq.\eqref{eq:jacobian}, and $\pmodel(x)$. We compare it to $\ptruth(x)$ and evaluate the heteroscedastic Gaussian log-likelihood loss
\begin{align}
    \loss_{\text{HNF}}
    = \frac{\left[ \pmodel(x)-\ptruth(x) \right]^{2}}{2\sigsyst^{2}(x)}
    + \log \sigsyst(x)  \eqp
    \label{eq:loss-het}
\end{align}
We have tested that it is sufficient to assume a Gaussian likelihood with a learned uncertainty $\sigsyst(x)$, but the heteroscedastic loss can also be extended to a Gaussian mixture model~\cite{Bollweg:2019skg,ATLAS:2024rpl}.

The default input to the MLP predicting $\sigsyst(x)$ is the concatenation of $x$ and $\pmodel(x)$. From our regression studies we expect the heteroscedastic NF (HNF) to learn a systematic uncertainty, defined as not vanishing in the limit of infinite training data of the same kind~\cite{Plehn:2022ftl,Bahl:2024gyt,Bahl:2025xvx,Dubey:2026cts}.

\subsubsection*{Accuracy and precision} 

We evaluate the HNF on a test dataset $\dtest$, and quantify local deviations between $\pmodel(x)$ and $\ptruth(x)$ using the relative accuracy,
\begin{align}
    \Deltatruth(x)
    = \frac{\pmodel(x)-\ptruth(x)}{\ptruth(x)} \eqp
    \label{eq:accuracy}
\end{align}
To assess the calibration of $\sigsyst(x)$ for a Gaussian likelihood, we define the pull~\cite{Plehn:2022ftl,Bahl:2024gyt, Bahl:2025xvx}
\begin{align}
    \pulltruth(x) = \frac{\pmodel(x)-\ptruth(x)}{\sigsyst(x)} \eqp
\end{align}

For a non-Gaussian likelihood~\cite{Bahl:2025xvx,Dubey:2026cts}, we can evaluate the empirical coverage for our test dataset and compute for what fraction of $\ptruth(x)$ the learned $\pmodel(x)$ lies within a confidence region,
\begin{align}
    \coverage = \XLangle
        \mathbb{1}\left(\pgauss(\ptruth(x) \,\vert\, \pmodel(x), \sigsyst(x)) > 1-\gamma\right)
    \XRangle_{\dtest} \eqc
\end{align}
written out for a Gaussian likelihood. The indicator function $\mathbb{1}$ is one if its condition is met and zero otherwise. The $p$-value for the true density $\ptruth(x)$ given the learned phase space dependent Gaussian probability distribution is
\begin{align}
    \pgauss(\ptruth(x) \,\vert\, \pmodel(x), \sigsyst(x))
    = 2\left[1-\Phi\left(\frac{\ptruth(x)-\pmodel(x)}{\sigsyst(x)}\right)\right] \eqc
    \label{eq:pvalue}
\end{align}
where $\Phi$ is the cumulative Gaussian distribution function. We show calibration curves in Appendix~\ref{app:results}.

\subsection{Bayesian normalizing flow}
\label{sec:bayesian}

Alternatively, we can learn the uncertainty on the density estimation by generative networks using Bayesian generative networks~\cite{Bellagente:2021yyh,Butter:2021csz,Butter:2023fov}. Instead of learning fixed values for the network parameters, a Bayesian network learns a posterior distribution for each parameter,
\begin{align}
    q_{\phi}(\theta) 
    \approx p(\theta\vert\dtrain) 
    &= \frac{p(\dtrain\vert \theta)p(\theta)}{p(\dtrain)} \notag \\
    &\equiv \frac{\pmodel(\dtrain)p(\theta)}{p(\dtrain)} \eqc
\end{align}
where $p(\theta)$ is a suitable prior and $p(\dtrain)$ is the training-irrelevant evidence. The success of the above training objective can be measured by 
\begin{align}
    \kl{q_{\phi}(\theta)}{p(\theta\vert\dtrain)} 
    &= - \Langle \log \pmodel(\dtrain) \Rangle_{q_{\phi}}
    + \kl{q_{\phi}(\theta)}{p(\theta)}
    + \text{const} \eqp
\end{align}
Choosing the parameter distributions and the prior both Gaussian, we can compute the KL-divergence analytically. It leads to the regularized likelihood loss of the Bayesian normalizing flow 
\begin{align}
    \loss_\text{BNF} = \frac{1}{\ntrain} \kl{q_{\phi}(\theta)}{p(\theta)} -
    \frac{1}{B} \sum_{b=1}^{B} \left\{ \log \pl(\overline{G}_{\theta}(x_{b}))
    + \log \left| \frac{\partial \overline{G}_{\theta}(x_{b})}{\partial x_{b}} \right| \right\} \eqp
    \label{eq:bnf-loss}
\end{align}
Here $\ntrain$ denotes the size of $\dtrain$ and $B$ the size of the mini-batches during training. 

For the generation, we sample the network parameters from the learned posterior and propagate them through the network to obtain the density estimate, such that
\begin{align}
    \theta &\sim q_{\phi}(\theta) \approx p(\theta\vert\dtrain) \notag \\
    x\vert\theta &\sim\pmodel(x)
    = \pl(\overline{G}_{\theta}(x)) \: 
    \left| \frac{\partial \overline{G}_{\theta}(x)}{\partial x} \right| \eqp
\end{align}
Repeated sampling of $\theta$ yields an ensemble of density estimates $\pmodel(x)$ at each point $x$, giving the statistical uncertainty on the density,
\begin{align}
    \sigstat^{2}(x)
    = \var_{\theta\vert\dtrain} \bigl[ \pmodel(x) \bigr] \eqp
\end{align}

\subsection{Classifier-reweighted target}
\label{sec:classifiertarget}

The main challenge of our HNF setup is the required explicit likelihood for the heteroscedastic loss. Most realistic applications do not come with an explicit likelihood. However, according to the Neyman--Pearson lemma, the ratio of the two likelihoods is the most powerful test statistic to distinguish the underlying hypotheses~\cite{Cranmer:2015bka,Metodiev:2017vrx}. This property of a learned classifier between generated and training data can be exploited to reweight the generated events towards the training data truth~\cite{Diefenbacher:2020rna,Butter:2021csz,Das:2023ktd}. The one disadvantage of this classifier reweighting is that the events need to be unweighted for experimental use.

We use the classifier weights relative to an approximate explicit probability distribution $\papprox(x)$. It has the same dimensionality and support as $\ptruth(x)$, and better approximations lead to better results. We then train a classifier $C(x)$ such that
\begin{align}    
    \frac{\ptruth(x)}{\papprox(x)} \approx \frac{C(x)}{1-C(x)} \eqp
\end{align}
We can then evaluate the combination
\begin{align}
    \papprox(x) \times \frac{C(x)}{1-C(x)} 
    \label{eq:ptarget}
\end{align}
as a precise, explicit approximation of $\ptruth(x)$ when training the generative HNF. Unlike in the standard approach, we use the classifier reweighting to train a generative network on an explicit density. This allows us to generate unweighted events and learn a local systematic uncertainty using a heteroscedastic loss. The only assumption we have to make is that the classifier is easier to train than the generative network, which is the basic assumption underlying classifier-reweighting and the technical reason for the success of classifier-based inference~\cite{Brehmer:2018eca} and unfolding~\cite{Andreassen:2019cjw}. 

\clearpage
\section{Toy model}
\label{sec:toy}

\begin{figure}[b!]
    \includegraphics[width=0.330\linewidth]{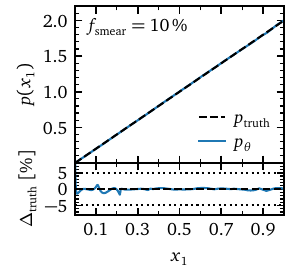}\hfill
    \includegraphics[width=0.330\linewidth]{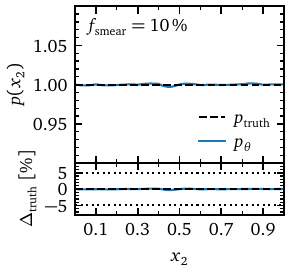}\hfill
    \includegraphics[width=0.330\linewidth]{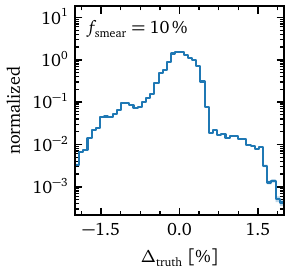}
	\caption{Marginalized true and learned HNF densities of the toy model with $10\,\%$ injected noise (left and center), evaluated on the test grid. Right: relative accuracy.}
	\label{fig:toy_density_fsmear10}
\end{figure}

We first study a two-dimensional toy model for which we know the density explicitly. It describes a wedge ramp~\cite{Bellagente:2021yyh}, linear in one direction and flat in the other,
\begin{align}
    p(x_1, x_2) = 2x_1 
    \qquad\mwith\quad 
    x_{1,2} \in [0,1] \eqp
    \label{eq:wedge}
\end{align}
We implement all networks using \pytorch~\cite{Paszke:2019xhz} and train them with the \adam optimizer~\cite{Kingma:2014vow} and its default parameters. For the NFs, we use the implementation provided by \madnis~\cite{Heimel:2023ngj}. We compose each NF of five rational-quadratic spline coupling blocks~\cite{Durkan:2019nsq} and we parameterize each block with fully-connected subnetworks. Each subnetwork contains a single hidden layer of width $32$. 

The MLP predicting $\sigsyst(x)$ consists of four hidden layers with $64$ neurons each. All networks use ReLU activations. The default input to the MLP is the concatenation of $x$ and $\pmodel(x)$. We train the NFs for $100$k iterations with a batch size of $512$ and a learning rate of $10^{-5}$ for the HNF and $10^{-3}$ for the BNF. For the HNF we train the NF and the MLP jointly. We evaluate the networks on a fixed test grid $\dtest$ of $500 \times 500$ points and list all hyperparameters in \cref{tab:setup}.

\subsection{Systematics from injected noise}
\label{sec:toy_noise}

\begin{figure}[t]
    \includegraphics[width=0.495\linewidth]{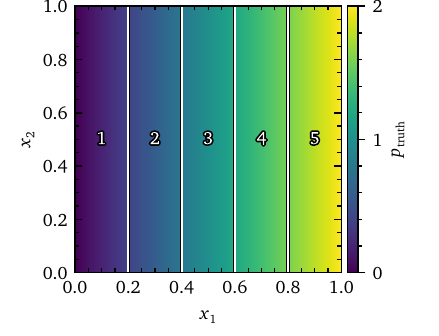}\hfill
    \includegraphics[width=0.495\linewidth]{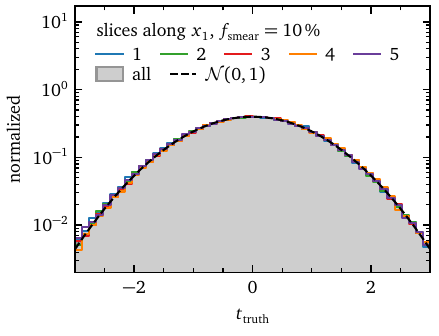}
	\caption{HNF pull distributions for the toy model with $10\,\%$ injected noise in phase space slices, evaluated on the test grid. The gray bulk in the right panel gives the pull over the full phase space.}
	\label{fig:toy_pull_fsmear10}
\end{figure}

\begin{figure}[b!]
    \centering
    \includegraphics[width=0.55\linewidth]{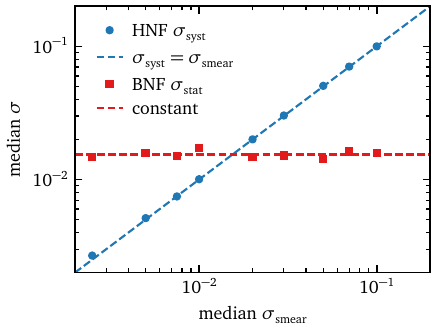}
	\caption{Learned HNF and BNF uncertainties and their expected scaling as a function of the injected noise for the toy model. Each point is the median from the test grid. The median $\sigsyst(x)$ for $\noisefrac = 0\,\%$ at $\num{7.56e-5}$ is not shown.}
	\label{fig:toy_noise}
\end{figure}

Our toy density is exact. To examine how well the HNF can learn a systematic effect, we inject artificial relative Gaussian noise into the explicit training density~\cite{Bahl:2024gyt}
\begin{align}
    \psmear(x) \sim \gauss(\ptruth(x), \sigsmear(x))
    \qquad\mwith\quad
    \sigsmear(x) = \noisefrac \times \ptruth(x) \eqp
    \label{eq:noiseinject}
\end{align}
We vary the noise level as
\begin{align}
    \noisefrac = \left\{ 0.25, 0.5, 0.75, 1, 2, 3, 5, 7, 10 \right\}\,\% \eqp
    \label{eq:fsmear}
\end{align}
A caveat is that for large noise levels we remove negative likelihoods and induce asymmetric pulls. In \cref{fig:toy_density_fsmear10} we first look at the marginal distributions of our wedge ramp with $10\,\%$ additional noise. We find excellent agreement with the truth and a clear peak in the relative accuracy around $\Deltatruth(x) = 0$ that indicates no bias. The width of the accuracy peak is driven by the added noise.

We then check whether the learned systematics are calibrated correctly. For relative noise, we can test how well the HNF learns the noise locally. We slice the phase space as indicated in the left panel of \cref{fig:toy_pull_fsmear10}. In the right panel we show the pulls for each slice. The gray bulk shows the pull distribution over the full phase space for $\noisefrac = 10\,\%$. All pulls are perfect standard Gaussians. This confirms that the HNF learns Gaussian systematics globally and locally.

Next, we reduce the noise level according to Eq.\eqref{eq:fsmear}. For comparison, we also train a BNF on the same datasets. As its log-likelihood we choose the MSE, and we confirm in Appendix~\ref{app:results} that the BNF output is not affected by the prior. In \cref{fig:toy_noise} we show the scaling of the uncertainty medians extracted from $\dtest$ with the input noise. Since the absolute value of $\sigsmear$ is phase space dependent, the injected noise is quantified by its median. The uncertainty learned by the HNF follows the injected noise level, whereas the learned BNF uncertainty fails. This confirms that injected noise constitutes a systematic uncertainty, which is learned by the heteroscedastic flow. The Bayesian NF is sensitive to statistical uncertainties and hence does not capture noise.

\begin{figure}[t]
    \centering
    \includegraphics[width=0.495\linewidth]{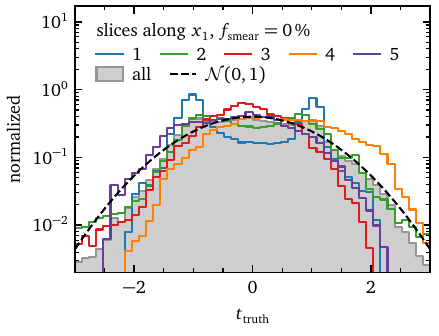}
	\caption{HNF pull distributions for the toy model with $\noisefrac = 0\,\%$ for the test grid slices defined in \cref{fig:toy_pull_fsmear10}. The gray bulk again shows the pull over the full phase space.}
	\label{fig:toy_pull_fsmear0}
\end{figure}

Finally, in \cref{fig:toy_pull_fsmear0} we show the pulls for the same slices without injected noise,
\begin{align}
    \noisefrac = 0\,\% \eqp
\end{align}
The corresponding pulls are non-Gaussian, and the low-density slice shows the most dramatic effects. This non-Gaussian shape is a known feature for low-dimensional phase spaces from amplitude regression~\cite{Bahl:2024gyt}. Only for uncorrelated higher-dimensional phase spaces the central limit theorem ensures Gaussian systematics. The symmetric peaks come from a persistent systematic offset $\pmodel(x) \neq \ptruth(x)$ that the learned uncertainty absorbs deterministically. On the positive side, we see that the HNF adapts the size of the learned systematics to the residual width, so each pull distribution has an approximate unit width.

\subsection{Systematics from learning rate}
\label{sec:toy_learning_rate}

\begin{figure}[b!]
	\centering
    \includegraphics[width=0.55\linewidth]{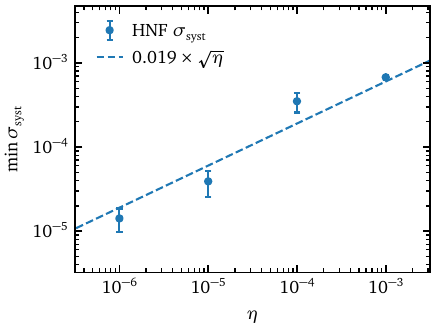}
	\caption{Minimum learned $\sigsyst(x)$ versus the learning rate $\eta$.}
    \label{fig:toy_min_sigma}
\end{figure}

\begin{figure}[t]
    \includegraphics[width=0.330\linewidth]{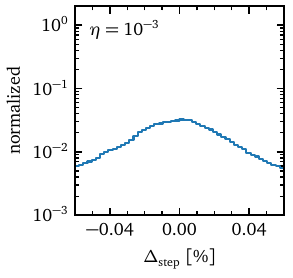}\hfill
    \includegraphics[width=0.330\linewidth]{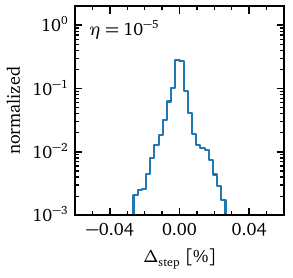}\hfill
    \includegraphics[width=0.330\linewidth]{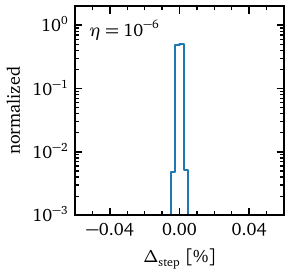}\\
    \includegraphics[width=0.330\linewidth]{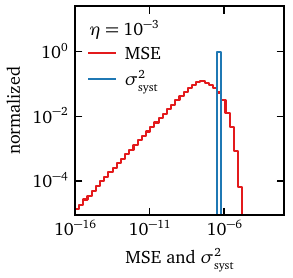}\hfill
    \includegraphics[width=0.330\linewidth]{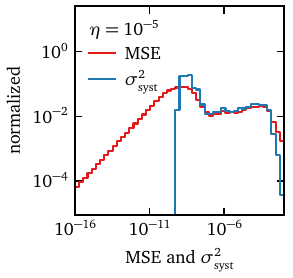}\hfill
    \includegraphics[width=0.330\linewidth]{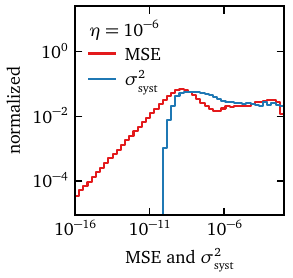}
	\caption{Upper: relative density change in optimization steps at the end of the training for $\eta = 10^{-3}, 10^{-5}, 10^{-6}$. Lower: $\text{MSE}(x) = [\pmodel(x) - \ptruth(x)]^2$ and $\sigsyst^{2}(x)$ versus the same learning rates.}
	\label{fig:toy_learning_rate}
\end{figure}

For a stochastic optimization process, like the network training, the weights converge to a distribution with a variance proportional to the learning rate $\eta$~\cite{mandt2017stochastic, malladi2022sdes}. This induces a variance of the network prediction, similar to a noisy target density, so we expect $\sigsyst(x)$ to reflect the learning rate. If the learning rate dominates the systematics and the variance of the network prediction is proportional to $\eta$, we expect the minimum of the learned uncertainty values to scale like 
\begin{align}
    \min_x \sigsyst(x) \propto \sqrt{\eta} \eqp
    \label{eq:noise_floor_lr}
\end{align}
To test this, we train the HNF with $\eta = 10^{-3}, \ldots, 10^{-6}$ for the $\pmodel(x)$ and $\sigsyst(x)$ to full convergence. In \cref{fig:toy_min_sigma} we show the minimum value of $\sigsyst(x)$ across $\dtest$ for each value of $\eta$ and fit it to Eq.\eqref{eq:noise_floor_lr}. The error bars correspond to $\pm 1\sigma$ over three independent trainings. The minimum value of $\sigsyst(x)$ follows this simple scaling within uncertainties.

To measure the size of the stochastic fluctuations from a finite learning rate, we track the learned density at consecutive optimization steps at the end of training,
\begin{align}
    \Delta_{\text{step},i}(x) = \frac{\pmodel(x)\big\vert_{\text{step}=i+1}-\pmodel(x)\big\vert_{\text{step}=i}}{\pmodel(x)\big\vert_{\text{step}=i}} 
    \qquad\mwith\quad x \in \dtest \eqp
    \label{eq:step_difference}
\end{align}
We collect the values over the last $25$ pairs of each training run. In the upper panels of \cref{fig:toy_learning_rate} we show the distributions of $\Delta_\text{step}(x)$ for three learning rates. All are centered at zero, so the learned density is not drifting anymore, and they sharpen with decreasing $\eta$. Their widths measure the relative density fluctuations between training steps, controlled by the learning rate.

We now turn to the effect of these fluctuations on the learned uncertainty. In the lower panels of \cref{fig:toy_learning_rate} we show the MSE between $\pmodel(x)$ and $\ptruth(x)$ together with the learned systematic uncertainty $\sigsyst^2(x)$, for three learning rates. The $\sigsyst(x)$ distributions shift to smaller values as $\eta$ decreases. Their lower edge follows the width of $\Delta_\text{step}(x)$ from the upper panels. This confirms that at large enough $\eta$ the minimal learned uncertainty is determined by the global step-to-step fluctuation of the network training.

The MSE distributions shown in the same panels shift in the same direction, but also change shape. For $\eta = 10^{-3}$ the MSE distribution has a single peak, as it is dominated by Gaussian optimization noise. As the learning rate decreases, the peak follows the decreasing optimization noise to smaller MSE values. In addition, a shoulder develops. We attribute it to a small persistent systematic offset $\pmodel(x) \neq \ptruth(x)$, just like the two peaks in \cref{fig:toy_pull_fsmear0}. 

\subsection{Limited training statistics}
\label{sec:toy_statistics}

\begin{figure}[t]
    \centering
    \includegraphics[width=0.55\linewidth]{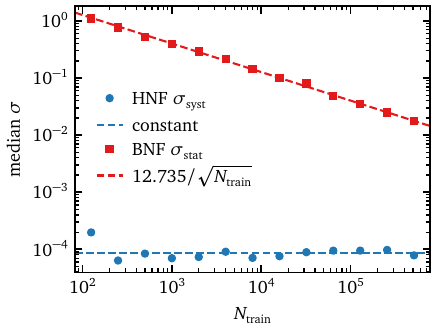}
	\caption{Learned HNF and BNF uncertainties as a function of the training dataset size $\ntrain$. Each point is the median from the test grid.}
    \label{fig:toy_statistics}
\end{figure}

Next, we study how the learned uncertainties of the HNF and BNF reflect the size of the training dataset,
\begin{align}
    \ntrain \in \{ 125, 250, 500, 1\text{k}, 2\text{k}, 4\text{k}, 8\text{k}, 16\text{k}, 32\text{k}, 64\text{k}, 128\text{k}, 256\text{k}, 512\text{k}\} \eqp
\end{align}
In \cref{fig:toy_statistics} we show the two learned uncertainties as a function of $\ntrain$. Each point is the median extracted from $\dtest$. As expected, the median of $\sigsyst(x)$ is constant across the range of $\ntrain$, while the median $\sigstat(x)$ learned by the BNF follows the expected scaling
\begin{align}
    \sigstat(x) \propto \frac{1}{\sqrt{\ntrain}} \eqp 
\end{align}
Indeed, the HNF and BNF capture different sources of uncertainty. The BNF learns the statistical limitations from the size of the training dataset while the HNF learns stochastic or deterministic systematics from the data, network, or training. Referring to them as data-driven or network-driven clearly makes no sense.

\subsection{Classifier-reweighted target}
\label{sec:toy_classifier}

\begin{figure}[t]
    \includegraphics[width=0.495\linewidth]{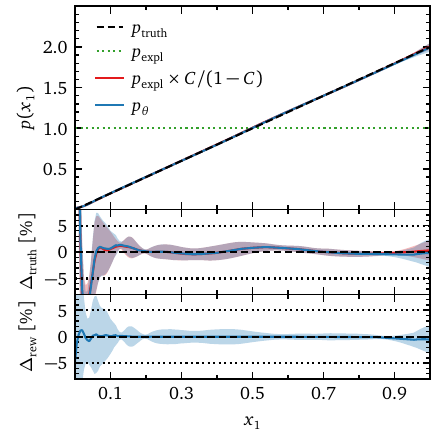}\hfill
    \includegraphics[width=0.495\linewidth]{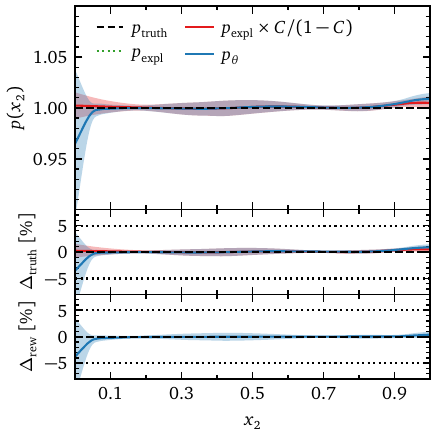}
	\caption{Relevant marginalized densities of the toy model on the test grid and the relative accuracies with respect to the truth and the classifier-reweighted density. The error bands correspond to $\pm 1\sigma$ over five independent trainings of the set of densities.}
    \label{fig:clf_marginals}
\end{figure}

From \cref{sec:classifiertarget} we know how to reconstruct $\ptruth(x)$ from an explicit density combined with a learned classifier in cases where we do not know $\ptruth(x)$ explicitly. The simple wedge ramp can be reconstructed from a simple uniform distribution,
\begin{align}
    \papprox(x_1, x_2) = 1
    \qquad\mwith\quad
    x_{1}, x_{2} \in [0,1] \eqp
    \label{eq:papprox_toy}
\end{align}
We train a classifier between $\papprox(x)$ and $\ptruth(x)$ on $25$k points, reweight $\papprox(x)$ according to Eq.\eqref{eq:ptarget}, and use this explicit combination to train the HNF. In \cref{fig:clf_marginals} we show the marginal distributions for
\begin{enumerate}
    \item the uniform $\papprox(x)$;
    \item the reweighted $\papprox(x) \times C(x)/(1-C(x))$;
    \item the learned $\pmodel(x)$; and 
    \item the truth $\ptruth(x)$. 
\end{enumerate}
In the sub-panels we show the relative accuracy with respect to $\ptruth(x)$, as defined in Eq.\eqref{eq:accuracy} and the relative accuracy with respect to the explicit classifier-reweighted density,
\begin{align}
    \Deltarew(x) = \frac{\pmodel(x) - \papprox(x)\dfrac{C(x)}{1-C(x)}}{\papprox(x)\dfrac{C(x)}{1-C(x)}} \eqp
\end{align}
The reweighted $\papprox(x) \times C(x)/(1-C(x))$ approximates $\ptruth(x)$ very well, and $\pmodel(x)$ follows this training target essentially perfectly.

\begin{figure}[b]
    \includegraphics[width=0.495\linewidth]{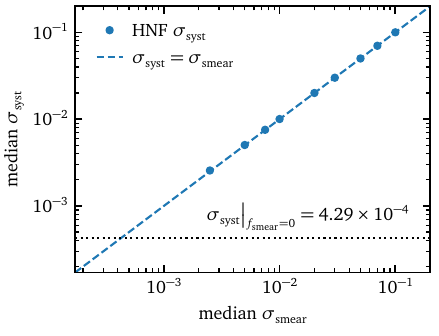}\hfill
    \includegraphics[width=0.495\linewidth]{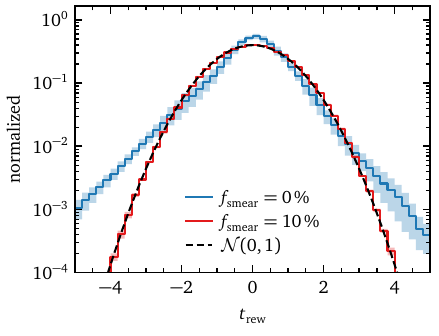}
	\caption{Left: HNF uncertainty as a function of the injected noise using classifier-reweighting. Each point is the median from the test grid. Right: pulls for $0\,\%$ and $10\,\%$ injected noise. Error bands are $\pm 1\sigma$ over five independent trainings.}
    \label{fig:clf_noise}
\end{figure}

As a last test, we evaluate the calibration of the uncertainty learned by the HNF using the classifier-reweighted density, again injecting Gaussian noise as described in Eq.\eqref{eq:noiseinject}. In the left panel of \cref{fig:clf_noise} we find excellent calibration of the learned systematic uncertainty. We indicate the negligible learned systematics for $\noisefrac = 0\,\%$. In the right panel we show the pull with respect to the explicit classifier-reweighted density,
\begin{align}
    \pullrew(x) = \frac{\pmodel(x) - \papprox(x)\dfrac{C(x)}{1-C(x)}}{\sigsyst(x)} \eqc
\end{align}
for the $0\,\%$ and $10\,\%$ noise levels. Their behavior aligns with the HNF results using $\ptruth(x)$ in \cref{sec:toy_noise}; for $\noisefrac = 10\,\%$ Gaussian noise, the pull follows a unit Gaussian, while without Gaussian noise, the systematic pull has much larger tails, but an order-one width.

\clearpage
\section{Top pair production}
\label{sec:physics}

As a physics benchmark, we apply our generative uncertainty quantification to top pair production with fully leptonic decays,
\begin{align}
    gg \;\to\; t\bar{t}
    \;\to\; (be^{+}\nu_{e})(\bar{b}e^{-}\bar{\nu}_{e}) \eqp
\end{align}
For simplicity, we only account for incoming gluons at LO and use MadAgents~\cite{Plehn:2026gxv} for the event generation with \madgraph~\cite{Alwall:2014hca}. We generate $100$k events or phase space points, split into a joint training and validation set with $80$k points and an evaluation dataset with $20$k points. 

Limiting ourselves to incoming gluons allows us to extract the truth amplitude and likelihood from \madgraph over the $16$-dimensional phase space basis
\begin{center}
    \begin{tabular}{@{\hspace{2pt}}lll@{\hspace{2pt}}}
        \toprule
        $1$\,--\,$4$    & $t\bar{t}$ production
        & $M_{tt} \,,\ y \,,\ \cos\theta^{*} \,,\ \phi^{*}$ \\
        $5$\,--\,$8$    & virtualities
        & $m_t^2 \,,\ m_{\bar t}^2 \,,\ m_{W^+}^2 \,,\ m_{W^-}^2$  \\
        $9$\,--\,$12$   & $t$ decay
        & $\cos\theta_{W^+} \,,\ \phi_{W^+} \,,\ \cos\theta_{\ell^+} \,,\ \phi_{\ell^+}$ \\
        $13$\,--\,$16$  & $\bar{t}$ decay
        & $\cos\theta_{W^-} \,,\ \phi_{W^-} \,,\ \cos\theta_{\ell^-} \,,\ \phi_{\ell^-}$ \\
        \bottomrule
    \end{tabular}
\end{center}
This likelihood only serves as the truth in the generative uncertainty evaluation, the actual uncertainty quantification does not need it. For illustration purposes, we also show the marginal truth distributions. For observables like the $\phi$-angles, we know their analytic form, for other observables, we show a smooth fit function. Again, these marginals are not needed for the actual uncertainty quantification.

\subsubsection*{Heteroscedastic flow}

We start again with an approximate explicit phase space distribution and refine it through classifier reweighting. Instead of the flat $\papprox(x)$ for the toy model, Eq.\eqref{eq:papprox_toy}, we follow two approaches to define $\papprox(x)$:
\begin{itemize}
    \item statistics-motivated, we train $\papprox(x)$ as a generative network on a smaller dataset and for only $100$ optimization steps, which starts to lack accuracy in all kinematic tails. Specifically, we use half of the full \madgraph training data.
    \item theory-motivated, we assume a factorized phase space where we ignore correlations between the phase space angles and replace each of them with its uniform marginal. We train $\papprox(x)$ only for $200$ optimization steps on the six non-angular phase space directions and $\cos\theta^*$, using half of our \madgraph training dataset.
\end{itemize}
The uncertainty-aware HNF training then follows the same steps as in \cref{sec:toy}
\begin{enumerate}
    \item Train $\papprox(x)$ using the statistical or theory approximation;
    \item Train the classifier $C(x)$ between samples from $\papprox(x)$ and the \madgraph training data;
    \item Reweight $\papprox(x)$ with $C(x)/(1-C(x))$;
    \item Train the HNF on the reweighted samples drawn from $\papprox(x)$.
\end{enumerate}
As the $t\bar{t}$ likelihood varies over several orders of magnitude, the network encodes the log-likelihood and the loss in Eq.\eqref{eq:loss-het} reads 
\begin{align}
    \loss_\text{HNF}
    = \frac{\left[ \log \pmodel(x)- \log\ptruth(x) \right]^{2}}{2\sigsyst^{2}(x)}
    + \log \sigsyst(x)  \eqc
    \label{eq:loss-het-log}
\end{align}
and similarly for the BNF loss in Eq.\eqref{eq:bnf-loss}. To keep the HNF training from overfitting, we continuously re-sample its training data from $\papprox(x)$ until it begins to converge. Then, we bootstrap by adding training points sampled from the HNF, as they describe $\ptruth(x)$ better~\cite{Butter:2024zbd}.

\begin{figure}[t]
    \includegraphics[page=01, width=0.495\linewidth]{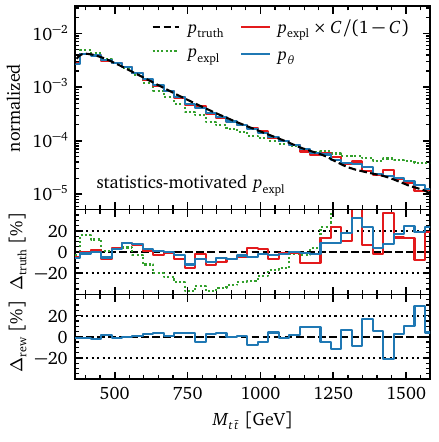}\hfill
    \includegraphics[page=02, width=0.495\linewidth]{figures/ttbar/ttbar_stat_marginals.pdf}\\
    \includegraphics[page=03, width=0.495\linewidth]{figures/ttbar/ttbar_stat_marginals.pdf}\hfill
    \includegraphics[page=11, width=0.495\linewidth]{figures/ttbar/ttbar_stat_marginals.pdf}
    \caption{Representative marginal distributions for the $t\bar{t}$ phase space, following all distributions needed to train $\pmodel(x)$ using the statistics-motivated $\papprox(x)$.}
	\label{fig:ttbar_marginals_stat}
\end{figure}

In \cref{fig:ttbar_marginals_stat} we compare representative marginals for the statistics-motivated $\papprox(x)$, its classifier-reweighted counterpart, and the learned $p_\theta(x)$ by the HNF to the truth. The explicit approximation overshoots for very small $M_{t\bar{t}}$ and in the large-$M_{t\bar{t}}$ tail, but the classifier reweighting corrects this behavior almost entirely. This shows that in statistically limited phase space regions the classifier is more efficient than the normalizing flow. In the second sub-panel we compare the HNF training target to the learned $\pmodel(x)$ and see that the percent-level limitation in the reweighted $\papprox(x)$ is inherited by $\pmodel(x)$. The rapidity distribution shows the same pattern in the far forward and backward directions. For all other marginals, the learned $p_\theta(x)$ from the HNF agrees with the truth at the few-percent level. We emphasize that the accuracy of the generative network could be improved, even for the normalizing flow~\cite{Butter:2021csz}, but the accuracy is not the main focus of this study.

\begin{figure}[t]
    \includegraphics[page=01, width=0.495\linewidth]{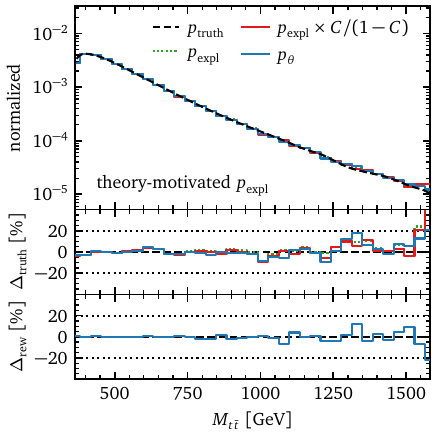}\hfill
    \includegraphics[page=11, width=0.495\linewidth]{figures/ttbar/ttbar_theo_marginals.pdf}
    \caption{Representative marginal distributions for the $t\bar{t}$ phase space, following all distributions needed to train $\pmodel(x)$ using the theory-motivated $\papprox(x)$.}
	\label{fig:ttbar_marginals_theo}
\end{figure}

For the theory-motivated $\papprox(x)$ we show the same comparison in \cref{fig:ttbar_marginals_theo}. Because the dimensionality of the learned phase space is smaller, the approximate $\papprox(x)$ is more accurate in the tails, for instance, at large $M_{t\bar{t}}$. In the right panel we show one of the angular marginal distributions. The initial $\papprox(x)$ is flat, with small statistical fluctuations. Reweighting it introduces the correct, slightly bent shape, which is then learned by the HNF accurately.

\subsection{Systematics}

\begin{figure}[t]
    \includegraphics[page=01, width=0.495\linewidth]{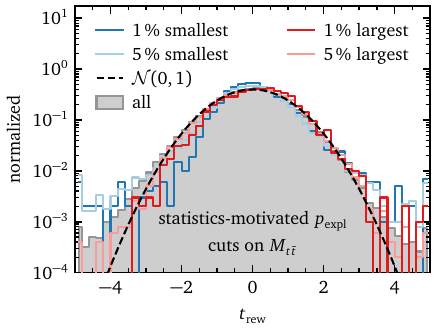}\hfill
    \includegraphics[page=11, width=0.495\linewidth]{figures/ttbar/ttbar_stat_pulls.pdf} \\
    \includegraphics[page=01, width=0.495\linewidth]{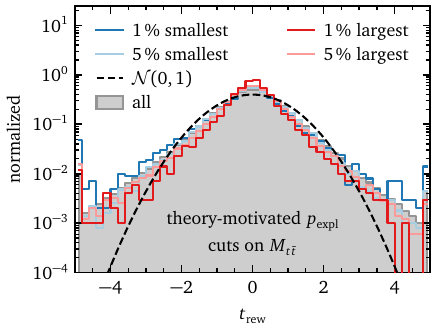}\hfill
    \includegraphics[page=11, width=0.495\linewidth]{figures/ttbar/ttbar_theo_pulls.pdf}
    \caption{Pulls for the quantile-based subsets for different observables, as defined in the text. We start from $\papprox(x)$ in the statistics-motivated (upper) and theory-motivated (lower) approximation.}
	\label{fig:ttbar_condcalib}
\end{figure}

Beyond the accuracy, we need to benchmark the reliability of the local $\sigsyst(x)$ from the HNF, trained on the reweighted density $\papprox(x) \times C(x)/(1-C(x))$. In the upper panels of \cref{fig:ttbar_condcalib} we show the Gaussian pulls for quantiles in two observables, $M_{t\bar{t}}$ and $\cos\theta_{\ell^+}$. In the upper left panel the light and dark red lines show the $5\,\%$ and $1\,\%$ largest $M_{t\bar{t}}$ values, moving into the tail of the distribution. In this regime the statistical limitations in the training data induces a Gaussian systematics, as indicated by the standard Gaussian shape of the pull. The dark and light blue lines show the $1\,\%$ and $5\,\%$ phase space points with the smallest $M_{t\bar{t}}$ values, close to the bulk of the distribution. Here, the systematics are not as Gaussian-dominated, explaining the larger tails of the centrally Gaussian-like pulls. For $\cos\theta_{\ell^+}$ the lowest and highest quantiles of an almost flat distribution are just slightly less Gaussian than the pull for the full dataset shown in gray.

In the lower panels of \cref{fig:ttbar_condcalib} we show the corresponding results for the theory-motivated $\papprox(x)$. For both observables, the global gray pull distributions are much less Gaussian, even though their distributions are centered around zero, uni-modal, with a typical unit width. Especially for the extreme $M_{t\bar{t}}$ quantiles, all of them are non-Gaussian and feature a sharp peak with large tails. As the HNF with a single $\sigsyst(x)$ implicitly assumes Gaussian systematics, we stick to the statistics-motivated construction of $\papprox(x)$ from here on, as it should and does lead to almost perfect Gaussian systematic pulls. We provide details on the non-Gaussian calibration in Appendix~\ref{app:results}.

\subsubsection*{Injected noise}

\begin{figure}[t]
    \includegraphics[width=0.495\linewidth]{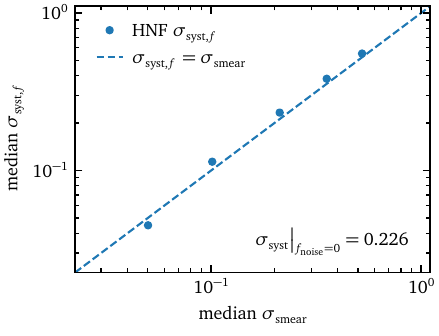}
    \hfill
    \includegraphics[width=0.495\linewidth]{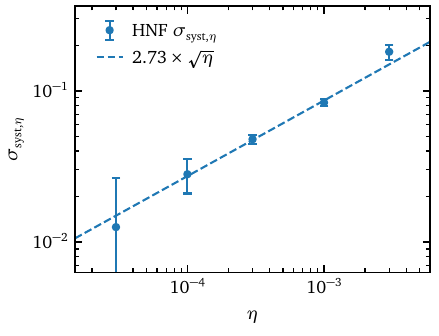}
    \caption{Left: median learned $\sigma_{\text{syst},f}(x)$ from the HNF versus the uncertainty from the injected noise, $\sigsmear$. Right: minimum learned $\sigma_{\text{syst},\eta}$ versus the learning rate $\eta$. The error bars correspond to $\pm 1\sigma$ over four independent trainings, excluding non-convergent runs.}
	\label{fig:deterministic_sigma_vs_target_noise}
\end{figure}

Like in \cref{sec:toy_noise} we inject artificial Gaussian noise into the HNF training. We adjust Eq.\eqref{eq:noiseinject} to a non-trivial and classifier-reweighted $\papprox(x)$ with a higher noise level,
\begin{align}
    \psmear(x) \sim \papprox(x) \; \frac{C(x)}{1-C(x)} \; \Big[ 1+ \noisefrac \times \normal (0,1) \Big] \notag \\
    \text{with} \qquad \noisefrac = \left\{ 5, 10, 20, 30, 40\right\} \,\% \eqp
\end{align}
Following Eq.\eqref{eq:loss-het-log}, the HNF learns the uncertainty on the log-density, so it does not absorb $\noisefrac$ itself, but the additive correction to the log density
\begin{align}
    \sigsmear^2 = \var \Bigg[ \log \Big[ 1 + \noisefrac \times \normal(0,1)  \Big] \Bigg] \eqp 
    \label{eq:noisewidth}
\end{align}
From above we know that the HNF learns a finite uncertainty without injected noise. Assuming uncorrelated Gaussian systematics we expect the combination to scale like 
\begin{align}
    \sigsyst^2(x) = \sigsyst^2(x) \Bigg|_{\noisefrac=0} + \sigma_{\text{syst}, f}^2(x) 
    \qquad\mwith\quad \sigma_{\text{syst}, f}(x) \simeq \sigsmear \eqp
\end{align}
We know $\sigsyst(x)$ without injected noise from above, which allows us to test the scaling of the learned and corrected $\sigma_{\text{syst},f}(x)$ with $\sigsmear$ in the left panel of \cref{fig:deterministic_sigma_vs_target_noise}. We see that the HNF learns the added Gaussian noise as a systematic uncertainty perfectly. 

\subsubsection*{Learning rate}

From \cref{sec:toy_learning_rate} we know the scaling of the minimal $\sigsyst(x)$ with a large learning rate $\eta$, as it induces noise into the network training. In analogy to the explicitly injected noise, we can test the scaling corresponding to \cref{fig:toy_min_sigma} for top pair production. Unlike for the injected noise, we do not know the baseline value for $\eta \to 0$, which means we first fit it using the assumed scaling of $\min \sigsyst(x)$ with $\eta$ and then subtract it 
\begin{align}
    (\min_x \sigsyst(x))^2 \approx (\min_x \sigsyst(x))^2 \Bigg|_{\eta \to 0} + \sigma_{\text{syst},\eta}^2 
    \qquad\mwith\quad \sigma_{\text{syst}, \eta} \propto \sqrt{\eta} \eqp
    \label{eq:noise_floor_lr_ttbar}
\end{align}
This relation turns into Eq.\eqref{eq:noise_floor_lr} when we assume that the baseline is negligible.

In the right panel of \cref{fig:deterministic_sigma_vs_target_noise} we show the learned and rescaled $\sigma_{\text{syst},\eta}$ versus the learning rate. Technically, we extract the minimum of the learned systematics as the lowest $0.1$ quantile. We start the training for each learning rate from a joint pre-trained network, to ensure convergence. Again, we observe a clear agreement between the expected and the actual scaling of the learned systematics, as the HNF captures noise in the network training reliably.

\subsection{Statistics}

The second source of uncertainty arises from statistical limitations of the training data and should be learned by the BNF. For the top pair events we vary the size of the training dataset as 
\begin{align}
    \ntrain \in \{ 2.5\text{k}, 10\text{k}, 20\text{k}, 40\text{k}, 80\text{k}, 160\text{k}, 320\text{k}, 640\text{k}, 1.25\text{M}, 2.4\text{M} \} \eqp
\end{align}
To ensure that we do not encounter saturation effects from the network sizes, we train two networks that differ only in their width, $256$ and $64$ hidden units per layer, corresponding to $6.4$M versus $1.3$M trainable parameters. 

\begin{figure}[t]
	\centering
    \includegraphics[width=0.55\linewidth]{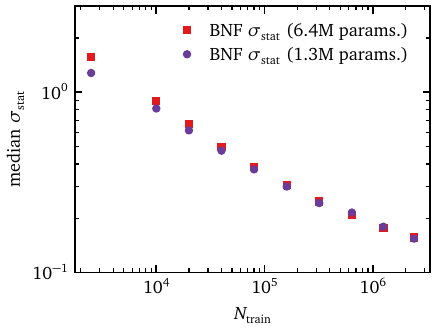}
	\caption{Learned BNF uncertainties as a function of the training dataset size $\ntrain$. Each point is the median from the test dataset.}
	\label{fig:ttbar_statistics}
\end{figure}

In \cref{fig:ttbar_statistics} we show the learned uncertainty from the BNF as a function of $\ntrain$. Each point is the median of the respective uncertainty extracted from $\dtest$. For a small training dataset the large network learns a larger uncertainty, which can be attributed to the known stability issues when training heavily over-parametrized Bayesian networks. The reason is that they can switch off the mean value encoded by a given parameter, but for an uninformative likelihood the width of the parameter distribution will move to the finite prior. For around $10^5$ training events both networks scale identically, and once the number of training events reaches the network size both networks deviate slightly from the power law scaling. This behavior is most likely a systematic uncertainty leaking into the BNF uncertainty budget, \ie, the attempt of the BNF to also accommodate systematics~\cite{Bahl:2024gyt}.

\subsection{Nuisance parameters}

\begin{figure}[b!]
    \includegraphics[width=0.495\linewidth]{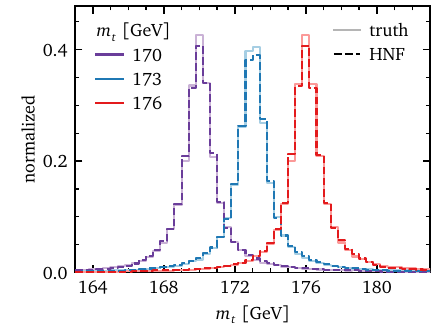}\hfill
    \includegraphics[width=0.495\linewidth]{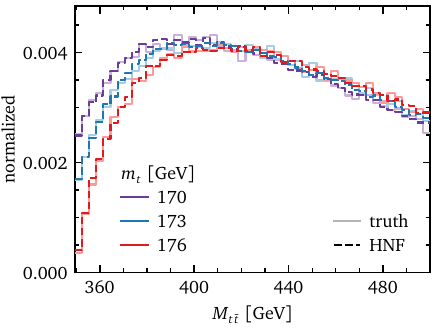}
	\caption{Kinematic distributions from the conditional HNF for three top mass values (dashed) and generated with \madgraph (solid).}
    \label{fig:mtt_conditional_marginal}
\end{figure}

The final question is how we can account for systematics known in terms of nuisance parameters. We know that it is possible to train a generative network conditional on a nuisance parameter varying the training data and then sample the generative networks and the nuisance parameter jointly~\cite{Butter:2021csz}. Here, we show that we can condition the generative network on a crucial parameter accurately. We vary the top mass as 
\begin{align}
    m_t \in \{ 167, 168, 169,\ldots,179\}\,\gev \eqp
\end{align}    
First, we train the statistics-motivated $\papprox(x)$ on events with different top masses, to ensure coverage. We then train a conditional classifier on the same kind of data, providing us with the training target for a conditional HNF
\begin{align}
    \papprox(x) \times \frac{C(x|m_t)}{1-C(x|m_t)} \eqp
\label{eq:papprox_cond}
\end{align}
In \cref{fig:mtt_conditional_marginal} we compare the kinematic distributions from the conditional HNF to events generated with \madgraph for the same top mass values and find perfect agreement for the top mass peaks themselves and for correlated observables. 

\begin{figure}[t]
    \includegraphics[width=0.495\linewidth]{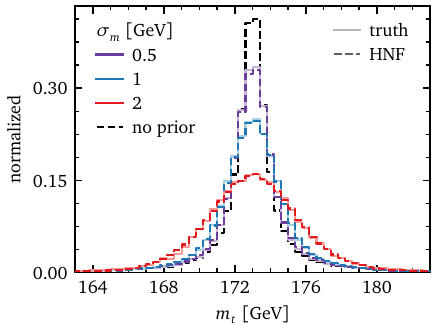}\hfill
    \includegraphics[width=0.495\linewidth]{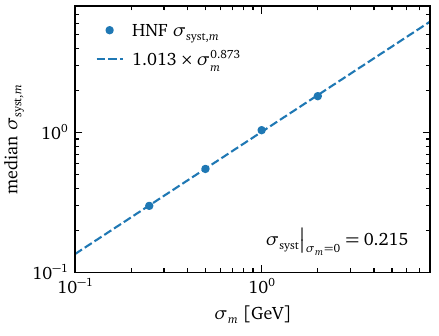}
	\caption{Left: reconstructed top mass from a conditional HNF with a Gaussian prior (dashed), compared to events from \madgraph with the same sampled top mass (solid). Right: learned and rescaled $\sigma_{\text{syst},m}$ versus the top mass prior width.}
    \label{fig:mt_nuisance}
\end{figure}

\subsubsection*{Prior sampling}

Next, we use the conditional HNF to sample from the top mass nuisance parameter. We choose a Gaussian prior to draw the top mass from
\begin{align}
    m_t \sim 173\,\gev + \normal(0, \sigma_m) 
    \qquad\mwith\quad \sigma_m \in \{ 0.25, 0.5, 1.0, 2.0\}~\gev \eqp
\end{align}
For the left panel of \cref{fig:mt_nuisance} we evaluate the conditional HNF with this Gaussian prior and compare it to a \madgraph generation with the same top mass sampling. Again, we find perfect agreement, which means we can use the conditional HNF to propagate nuisance parameters with known priors to all phase space observables.

Finally, we train the HNF on Eq.\eqref{eq:papprox_cond}, but without conditioning. This way, the HNF will absorb the varying top mass in the training target into a systematic uncertainty, 
\begin{align}
    \sigsyst^2 = \sigsyst^2 \Bigg|_{\sigma_m=0} + \sigma_{\text{syst}, m}^2 
    \qquad\mwith\quad \sigma_{\text{syst}, m} \propto \sigma_m^\kappa  \eqp
\end{align}
In the right panel of \cref{fig:mt_nuisance} we show the learned and rescaled uncertainty as a function of the prior width $\sigma_m$. Knowing the systematics for the nominal top mass we can extract the underlying relation between the width of the top mass prior and the corresponding systematics as 
\begin{align}
    \sigsyst^2 = \sigsyst^2 \Bigg|_{\sigma_m=0} + \left( 1.013 \times \sigma_m^{0.873} \right)^2 \eqp
\end{align}
Unlike for the injected noise and the learning rate, we cannot predict this form from the complex error propagation, so this scaling law constitutes a genuinely new scientific result.

\clearpage
\section{Outlook}

We have demonstrated how to obtain a comprehensive uncertainty estimate for generative networks. We first introduced a heteroscedastic normalizing flow to capture systematic uncertainties in the training of generative networks with known target densities. It captures systematic uncertainties from noisy data as well as systematic uncertainties from the network training (and eventually the network architecture). We explicitly showed that the learned systematic uncertainties are locally correctly calibrated. Statistical uncertainties, induced by a lack of training data, can be estimated with a Bayesian normalizing flow, and we have shown that they scale correctly with the size of the training dataset. The combination of HNF and BNF allows for a comprehensive uncertainty quantification of generative networks. 

While the HNF requires an explicit target density, most particle physics problems do not have access to the density. We have addressed this challenge by introducing a classifier-reweighted approximate generative network as an alternative target for the HNF training. This turns the unsupervised training of generative networks into a supervised problem for which we can estimate heteroscedastic systematics. As possible approximations we used a low-statistics training for a limited number of iterations and a training without taking into account correlations. Using the example of top pair production we showed that the combination of HNF and BNF learn the systematic and statistical uncertainties correctly, with the quantitative caveats of a realistic physics problem. Finally, we have shown that a conditional HNF can learn the systematic uncertainties encoded in a nuisance parameter and propagate them reliably to the entire phase space.
    
\subsection*{Code availability}

Our code is available in our joint GitHub at \url{https://github.com/heidelberg-hepml/}.

\section*{Acknowledgements}

We would like to thank Henning Bahl for many inspiring discussions. This research is supported through the KISS consortium (05D2022) funded by the German Federal Ministry of Education and Research BMBF in the ErUM-Data action plan and by the Deutsche Forschungsgemeinschaft (DFG, German Research Foundation) under grant 396021762 -- TRR 257 \textsl{Particle Physics Phenomenology after the Higgs Discovery}. Part of this work was supported by the Carl-Zeiss-Stiftung through the project \textsl{Model-Based AI: Physical Models and Deep Learning for Imaging and Cancer Treatment}. AB gratefully acknowledges the continuous support from LPNHE, CNRS/IN2P3, Sorbonne Université, and Université de Paris Cité. This project and SD are supported by the Baden-W\"urttemberg Stiftung. We acknowledge support by the European Union's Horizon Europe research and innovation programme under the Marie Sk\l{}odowska-Curie grant agreement No.~101168829, \textsl{Challenging AI with Challenges from Physics: How to solve fundamental problems in Physics by AI and vice versa} (AIPHY). 

\appendix

\clearpage
\section{Hyperparameters}
\label{app:hyperparameters}

\begin{table}[H]
	\centering
	\begin{small}
		\begin{tabular}{@{\hspace{2pt}}l|l@{\hspace{2pt}}}
			\toprule
            hyperparameter & HNF and BNF network \\
            \midrule
            network optimizer & \adam algorithm~\cite{Kingma:2014vow} with default parameters \\
                      & ($\beta_1=0.9$, $\beta_2=0.999$, $\epsilon=10^{-8}$, weight decay $\lambda=0$) \\
            learning rate $\eta$ & $10^{-5}$ HNF, $10^{-3}$ BNF (constant, no $\eta$ scheduler) \\
            batch size $B$ & $512$ \\
            \# training samples $\ntrain$ & $\num{300000}$ \\
            \# training iterations & $\num{100000}$ \\
            \midrule
            coupling block type & rational-quadratic splines (RQS)~\cite{Durkan:2019nsq} \\
            \# spline bins & $10$ \\
            \# coupling blocks & $5$ \\
            \# hidden layers per block & $1$ \\
            \# units per hidden layer & $32$ \\
            activation function & rectified linear unit (ReLU) \\
            \midrule
            \# trainable parameters & $\num{27676}$ HNF, $\num{21430}$ BNF \\
			\bottomrule
		\end{tabular}
	\end{small}
    \caption{Network architectures and training setup for our toy model in \cref{sec:toy}. For the network in the HNF setup that predicts $\sigsyst(x)$ we use an MLP consisting of four hidden layers with $64$ neurons each and ReLU activations. The input to the MLP is the concatenation of the phase space points $x$ and the NF-predicted density $\pmodel(x)$. For the HNF setup we train the NF and the MLP jointly with the same optimizer.}
	\label{tab:setup}
\end{table}

\begin{table}[H]
	\centering
	\begin{small}
		\begin{tabular}{@{\hspace{2pt}}l|l@{\hspace{2pt}}}
			\toprule
            hyperparameter & $t\bar{t}$ $\papprox(x)$ network \\
            \midrule
            network optimizer & \adam algorithm~\cite{Kingma:2014vow} with default parameters \\
                      & ($\beta_1=0.9$, $\beta_2=0.999$, $\epsilon=10^{-8}$, weight decay $\lambda=10^{-5}$) \\
            learning rate $\eta$ & $3\times 10^{-4}$ (one-cycle $\eta$ scheduler, warm-up fraction $0.1$) \\
            batch size $B$ & $\num{4096}$ \\
            \# training samples $\ntrain$ & $\num{80000}$ \\
            \# training iterations & \num{100} statistics-motivated, \num{200} theory-motivated \\
            \midrule
            coupling block type & rational-quadratic splines (RQS)~\cite{Durkan:2019nsq} \\
            \# input dimensions & \num{16} statistics-motivated, \num{9} theory-motivated \\
            \# spline bins & $32$ \\
            \# coupling blocks & $12$ \\
            \# hidden layers per block & $3$ \\
            \# units per hidden layer & $256$ \\
            activation function & rectified linear unit (ReLU) \\
            \midrule
            \# trainable parameters & $\num{1998384}$ \\
			\bottomrule
		\end{tabular}
	\end{small}
    \caption{Network architecture and training setup for the approximate phase space density
    $\papprox(x)$ in \cref{sec:physics}.}
	\label{tab:setup_papprox}
\end{table}

\begin{table}[H]
	\centering
	\begin{small}
		\begin{tabular}{@{\hspace{2pt}}l|l@{\hspace{2pt}}}
			\toprule
            hyperparameter & $t\bar{t}$ classifier network $C(x)$ \\
            \midrule
            network optimizer & \adam algorithm~\cite{Kingma:2014vow} with default parameters \\
                      & ($\beta_1=0.9$, $\beta_2=0.999$, $\epsilon=10^{-8}$, weight decay $\lambda=10^{-4}$) \\
            learning rate $\eta$ & $10^{-3}$ (one-cycle $\eta$ scheduler, warm-up fraction $0.1$) \\
            batch size $B$ & \num{4096}, balanced $50/50$ between the two classes \\
            \# training samples $\ntrain$ & \num{80000} \madgraph events, disjoint from $\papprox(x)$ \\
            \# training iterations & $\num{8000}$ \\
            \midrule
            network type & fully-connected MLP \\
            \# input features & $18$ ($16$ observables + $2$ rescaled) \\
            \# hidden layers & $5$ \\
            \# units per hidden layer & $\num{1024}$ \\
            activation function & rectified linear unit (ReLU) \\
            dropout rate & $0.4$ \\
            \midrule
            \# trainable parameters & $\num{3169281}$ per network, $\num{31692810}$ for the ensemble \\
			\bottomrule
		\end{tabular}
	\end{small}
    \caption{Network architecture and training setup for the classifier $C(x)$ in
    \cref{sec:physics}. The same setup is used for both $\papprox(x)$ variants. 
    $\papprox(x)$ samples are redrawn at every iteration, true  samples are augmented using
    the exact azimuthal symmetry of the process. We ensemble over ten independent classifier trainings. 
    The classifier weights are clipped to $w\in[1/5000,50]$ for HNF training stability}
	\label{tab:setup_classifier}
\end{table}

\begin{table}[H]
	\centering
	\begin{small}
		\begin{tabular}{@{\hspace{2pt}}l|l@{\hspace{2pt}}}
			\toprule
            hyperparameter & $t\bar{t}$ HNF network \\
            \midrule
            network optimizer & \adam algorithm~\cite{Kingma:2014vow} with default parameters \\
                      & ($\beta_1=0.9$, $\beta_2=0.999$, $\epsilon=10^{-8}$, weight decay $\lambda=0$) \\
            learning rate $\eta$ & $3\times 10^{-4}$ (one-cycle $\eta$ scheduler, warm-up fraction $0.1$) \\
            batch size $B$ & \num{8192} \\
            \# training samples $\ntrain$ & $\num{200000}$ points drawn from $\papprox(x)$ \\
            \# training iterations & $\num{15000}$ \\
            \midrule
            coupling block type & rational-quadratic splines (RQS)~\cite{Durkan:2019nsq} \\
            \# spline bins & $48$ \\
            \# coupling blocks & $20$ \\
            \# hidden layers per block & $3$ \\
            \# units per hidden layer & $192$ \\
            activation function & rectified linear unit (ReLU) \\
            \midrule
            \# trainable parameters & $\num{5253280}$ \\
			\bottomrule
		\end{tabular}
	\end{small}
    \caption{Network architecture and training setup for the heteroscedastic flow in
    \cref{sec:physics}. Training points are drawn fresh from $\papprox(x)$ at every iteration. As in
    \cref{tab:setup}, the MLP predicting $\sigsyst(x)$ is trained jointly with the flow using
    the same optimizer; here it consists of six hidden layers with $64$ neurons each
    ($\num{17857}$ parameters), taking as input the concatenation of $x$ and $\pmodel(x)$.}
	\label{tab:setup_hnf}
\end{table}
\begin{table}[H]
	\centering
	\begin{small}
		\begin{tabular}{@{\hspace{2pt}}l|l@{\hspace{2pt}}}
			\toprule
            hyperparamter & $t\bar{t}$ conditional classifier network $C(x|m_t)$ \\
            \midrule
            network optimizer & \adam algorithm~\cite{Kingma:2014vow} with default parameters \\
                      & ($\beta_1=0.9$, $\beta_2=0.999$, $\epsilon=10^{-8}$, weight decay $\lambda=10^{-4}$) \\
            learning rate $\eta$ & $10^{-3}$ (one-cycle $\eta$ scheduler, warm-up fraction $0.1$) \\
            batch size $B$ & $\num{4096}$, balanced $50/50$ between the two classes \\
            \# training samples $\ntrain$ & $\num{1300000}$ \madgraph events, disjoint from $\papprox(x)$ \\
            \# training iterations & $\num{8000}$ \\
            \midrule
            network type & fully-connected MLP \\
            \# input features & $17$ ($16 + m_t$) \\
            \# hidden layers & $5$ \\
            \# units per hidden layer & $512$ \\
            activation function & rectified linear unit (ReLU) \\
            \midrule
            \# trainable parameters & $\num{797697}$ \\
			\bottomrule
		\end{tabular}
	\end{small}
    \caption{Network architecture and training setup for the mass-conditional classifier
    $C(x|m_t)$ used for the nuisance-parameter study in \cref{sec:physics}. The top mass is
    appended to the feature vector. Its $\papprox(x)$ is trained on a pooled dataset of events
    spread equally over the $13$ integer masses $m_t = 167 \dots 179\,\gev$, so both classes cover
    the whole range, and the classifier sees the half of that pool the flow did not.
    No explicit augmentation is applied.}
	\label{tab:setup_classifier_cond}
\end{table}

\clearpage
\section{Supplementary results}
\label{app:results}

\subsection*{BNF prior dependence}

\begin{figure}[b!]
    \includegraphics[width=0.495\linewidth]{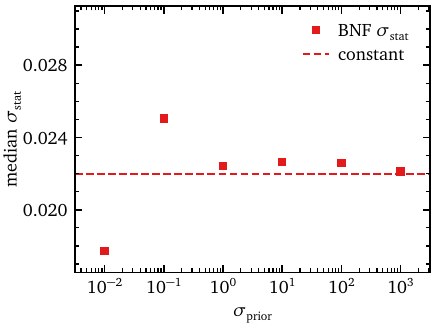}\hfill
    \includegraphics[width=0.495\linewidth]{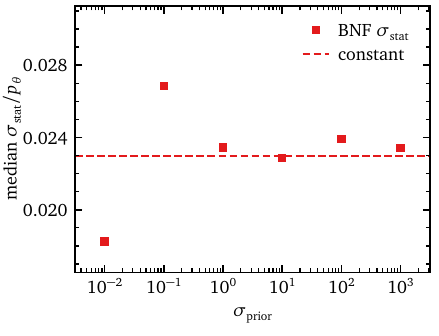}
	\caption{Absolute (left) and relative (right) BNF uncertainties as a function of the Gaussian prior width for the toy model.}
    \label{fig:toy_bnf_prior}
\end{figure}

As for any Bayesian setup, it is crucial that we confirm that the output of our BNF is not dominated by the prior. Our default prior is a Gaussian with $\mu_{\text{prior}} = 0$ and $\sigma_{\text{prior}} = 1$. To test the prior dependence in the toy model we stick to Gaussians with $\mu_{\text{prior}}=0$, but vary the width of the prior distribution over an extremely wide range,
\begin{align}
    \sigma_{\text{prior}} \in \{ 10^{-2}, 10^{-1}, 10^{0}, 10^{1}, 10^{2}, 10^{3} \} \eqp
\end{align}
We show the results for this test in \cref{fig:toy_bnf_prior}. The learned $\sigstat(x)$ and corresponding relative uncertainty $\sigstat(x)/\pmodel(x)$ are approximately flat across six orders of magnitude in $\sigma_\text{prior}$.

\subsection*{Top pair quantile calibration}

As a complementary check to the pull distributions in \cref{fig:ttbar_condcalib}, we further examine the calibration curves, as defined in Eq.\eqref{eq:pvalue}, using $\papprox(x) \times C(x)/(1-C(x))$ instead of $\ptruth(x)$ and the same quantile regions as defined in \cref{sec:flow_het}.

\begin{figure}[t!]
    \includegraphics[page=01, width=0.495\linewidth]{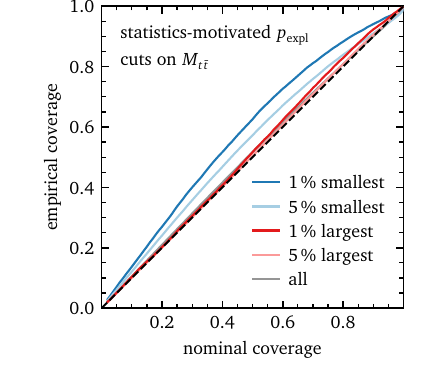}\hfill
    \includegraphics[page=02, width=0.495\linewidth]{figures/ttbar/ttbar_stat_coverage.pdf}\\
    \includegraphics[page=01, width=0.495\linewidth]{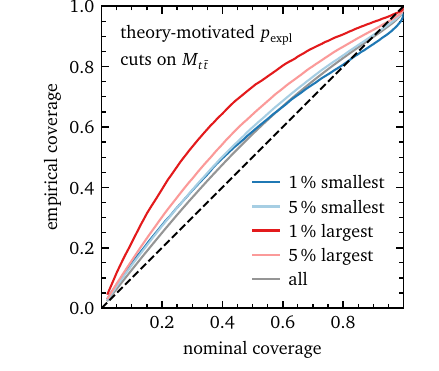}\hfill
    \includegraphics[page=02, width=0.495\linewidth]{figures/ttbar/ttbar_theo_coverage.pdf}
	\caption{Calibration curves for the quantile-based subsets for different observables, as defined in \cref{fig:ttbar_condcalib}. We start from $\papprox(x)$ in the statistics-motivated (upper) and theory-motivated (lower) approximation.}
    \label{fig:ttbar_calibration_curves}
\end{figure}

\Cref{fig:ttbar_calibration_curves} shows the resulting calibration curves. The top row shows the curves we get using the statistics-motivated $\papprox(x)$, while the bottom row shows the results of the theory-motivated $\papprox(x)$. The left-hand column shows calibration curves for the $M_{t\bar{t}}$ quantiles, while the right-hand column shows curves for the $\cos\theta_{\ell^+}$ quantiles.
We see that the curves which most strongly deviate from the diagonal are those with the most notably asymmetric pulls in \cref{fig:ttbar_condcalib}. Specifically, the $M_{t\bar{t}}$ quantiles combined with the theory-motivated $\papprox(x)$ show significant deviation. For the statistics-motivated $\papprox(x)$, the lower quantiles (blue lines) show the largest miscalibration, and these are again the quantiles that produce the most asymmetric pulls. 
In contrast, the quantile split in $\cos\theta_{\ell^+}$ shows very well calibrated curves for both $\papprox(x)$, which is congruent with the more symmetric pulls for quantiles in $\cos\theta_{\ell^+}$. 
Notably, the presence of non-Gaussian tails in the pull distribution has little impact on the calibration curve, as visible in the $\cos\theta_{\ell^+}$ case. 

\clearpage
\bibliographystyle{tepml}
\bibliography{tilman,refs,lorenz}

\end{document}

%% file: incl_settings.tex
\usepackage[utf8]{inputenc} 
\usepackage[T1]{fontenc} 	
\usepackage[english]{babel} 

\usepackage[bitstream-charter]{mathdesign}
\usepackage{geometry} 		
\usepackage{amsmath} 		
\usepackage{mathtools} 		
\usepackage{float} 			
\usepackage{graphicx} 		
\usepackage{tabularx} 		
\usepackage{booktabs} 		
\usepackage{color, xcolor} 	
\usepackage{pdfpages} 		
\usepackage{extarrows} 		
\usepackage{multirow} 		
\usepackage{multicol} 		
\usepackage{caption} 		
\usepackage{comment}
\usepackage{subcaption} 	
\usepackage{enumitem} 		
\usepackage{xspace} 		
\usepackage{stackrel} 		
\usepackage{tikz} 			
\usetikzlibrary{calc}
\usepackage{braket} 		
\usepackage{bm} 			
\usepackage{tensor} 		
\usepackage{slashed} 		
\usepackage{siunitx} 
\usepackage{lastpage} 		
\usepackage{cite} 			
\usepackage[normalem]{ulem} 
\usepackage{fontawesome} 	
\usepackage{tocloft} 		
\usepackage{titlesec} 		
\usepackage{doi} 			
\usepackage[most]{tcolorbox} 					
\usepackage[capitalize]{cleveref} 	
\usepackage[nottoc, notlot, notlof]{tocbibind} 	
\usepackage[ruled, vlined]{algorithm2e} 		
\usepackage{makecell}
\usepackage{pifont}

\hypersetup{
	pdftitle={Know What You Dont Flow},
	pdfauthor={Anja Butter, Sascha Diefenbacher, Tilman Plehn, and Lorenz Vogel},
    linktoc=section,
	breaklinks=true,			
	colorlinks=true, 			
	linkcolor={red!50!black}, 	
	citecolor={blue!50!black}, 	
	urlcolor={blue!80!black} 	
} 

\DeclareSymbolFont{usualmathcal}{OMS}{cmsy}{m}{n}
\DeclareSymbolFontAlphabet{\mathcal}{usualmathcal}

\SetArgSty{textnormal}
\SetKwComment{Comment}{{\small\#}~}{}
\SetCommentSty{mycommfont}

\setitemize{itemsep=2pt, parsep=0pt} 				
\setenumerate{itemsep=2pt, parsep=0pt} 				
\crefname{section}{Section}{Sections}
\crefname{equation}{Eq.}{Eqs.}
\crefname{figure}{Figure}{Figures}
\crefname{table}{Table}{Tables}
\Crefname{section}{Section}{Sections}
\Crefname{equation}{Equation}{Equations}
\Crefname{figure}{Figure}{Figures}
\Crefname{table}{Table}{Tables}

%% file: incl_shortcuts.tex
\newcommand{\ie}{i.e.\@\xspace} 	

\newcommand{\eqc}{\;\text{,}} 		
\newcommand{\eqp}{\;\text{.}} 		

\newcommand{\pl}{p_\text{latent}}

\newcommand{\Langle}{\bigl\langle}
\newcommand{\Rangle}{\bigr\rangle}
\newcommand{\XLangle}{\Bigl\langle}
\newcommand{\XRangle}{\Bigr\rangle}

\newcommand{\mwith}{\text{with}}

\newcommand{\mor}{\text{or}}

\newcommand{\var}{\operatorname{Var}} 		
\newcommand{\normal}{\mathcal{N}} 			

\newcommand{\loss}{\mathcal{L}} 	

\newcommand{\gauss}{\mathcal{N}} 	

\newcommand{\kl}[2]{D_{\text{KL}}[#1,#2]}

\newcommand{\pytorch}{\texttt{PyTorch}\xspace}

\newcommand{\madnis}{\texttt{MadNIS}\xspace}

\newcommand{\madgraph}{\textsc{MadGraph}\xspace}

\newcommand{\sherpa}{\textsc{Sherpa}\xspace}

\newcommand{\adam}{\texttt{Adam}\xspace}

\newcommand{\arXiv}[2][]{%
	\ifthenelse{\equal{#1}{}}%
	{\href{http://arxiv.org/abs/#2}{arXiv:#2}}%
	{\href{http://arxiv.org/abs/#2}{arXiv:#2~[#1]}}}

\newcommand{\gev}{\text{GeV}}

\def\slashchar#1{\setbox0=\hbox{$#1$}           
   \dimen0=\wd0                                 
   \setbox1=\hbox{/} \dimen1=\wd1               
   \ifdim\dimen0>\dimen1                        
      \rlap{\hbox to \dimen0{\hfil/\hfil}}      
      #1                                        
   \else                                        
      \rlap{\hbox to \dimen1{\hfil$#1$\hfil}}   
      /                                         
   \fi}

\newcommand{\tikznode}[2]{%
\ifmmode%
\tikz[remember picture,baseline=(#1.base),inner sep=0pt] \node (#1) {$#2$};%
\else
\tikz[remember picture,baseline=(#1.base),inner sep=0pt] \node (#1) {#2};%
\fi}

\def\mathswitchr#1{\relax\ifmmode{\text{#1}}\else$\text{#1}$\xspace\fi}
\def\mathswitch#1{\relax\ifmmode#1\else$#1$\xspace\fi}

\newcommand{\dtrain}{\mathcal{D}_{\text{train}}}
\newcommand{\dtest}{\mathcal{D}_{\text{test}}}

\newcommand{\ntrain}{N_{\text{train}}}

\newcommand{\pmodel}{p_{\theta}}               
\newcommand{\ptruth}{p_{\text{truth}}}                
\newcommand{\psmear}{p_\text{smear}}         

\newcommand{\sigsyst}{\sigma_{\text{syst}}}          
\newcommand{\sigstat}{\sigma_{\text{stat}}}    
\newcommand{\sigsmear}{\sigma_{\text{smear}}}  
\newcommand{\noisefrac}{f_{\text{smear}}}      

\newcommand{\coverage}{c_{\gamma}}             
\newcommand{\pgauss}{\mathcal{P}_{\text{Gauss}}}         

\newcommand{\papprox}{p_{\text{expl}}}      

\newcommand{\Deltatruth}{\Delta_{\text{truth}}}
\newcommand{\Deltarew}{\Delta_{\text{rew}}}
\newcommand{\pulltruth}{t_{\text{truth}}}
\newcommand{\pullrew}{t_{\text{rew}}}